# Numerical Studies of the Evolutionary Rate of Mutant Allele at Duplicate Loci

Yupeng Cun

Bonn-Aachen International Center for IT, Algorithmic Bioinformatics, Dahlmannstrße 2, 53113, Bonn, Germany

Contact mail: cun@bit.uni-bonn.de

**Abstract**

Gene duplication is one of the major primary driving evolutionary forces. Gene and genome duplication have increased the number of genes in the genome of species and the complexity of genome architecture. Several population genetics models had been used to study the evolutionary fate of duplicate gene. In this paper we will introduce a stochastic difference equation model to investigate, how evolutionary forces act during the fixation of a mutant allele at duplicate loci. We study the fixation time of a mutant allele at duplicate loci under a double null recessive model (DNR) and a haploinsufficient model (HI). We also look at how selection coefficients together with another evolutionary force influence the fixation frequency of a mutant allele at duplicate loci. Our results suggest that selection plays a role in the evolutionary fate of duplicate genes, and that tight linkage helps the mutant allele to stay at duplicate loci. Our theoretical simulations agree well with existing genomics data. This specifically includes the fact that selection, rather than drift, plays an important role in the establishment of duplicate loci, and that recombination occur with strong selection would help the mutant allele to stay at duplicate loci.

# 1. INTRODUCTION

Gene duplications are one of the major driving forces in the evolution of genomes (Ohno, 1970; Zhang, 2003; Gu et al., 2003). Duplicate genes are believed to be a major source for the establishment of new gene functions (Van de Peer et al., 2001; Blanc et al., 2003; Moore & Purugganan, 2003; Moore & Purugganan, 2005; Roth et al., 2007; Innan & Kondrashov, 2010). As the emergence of genome sequences data, there are many studies about the mechanism and patterns of gene/genome duplication (Gu et al., 2003; Moore & Purugganan, 2003; Force et al., 1999; Lynch & Conery 2000; Bowers et al., 2003; Blanc & Wolfe, 2004; Clauss & Mitchell-Olds, 2004; Semon & Wolfe, 2008; Kleinjan et al., 2008). Since the remarkable book by Ohno (1970), the reigning paradigm regarding the fate of duplicate genes is that one of the duplicates is either lost (pseudogenization) or gains a new function (neofunctionalization). According the classical population genetics theory, lethal mutations have larger probability than advantageous mutations. It also presumed that most duplicate genes are lost and only a few remain as new genes.

Fisher presented the first population genetics model of the fate of duplicate genes (Fisher, 1935). Afterwards there were several classical works on strict gene silencing models (Nei, 1970; Bailey et al., 1978; Takahata & Maruyama, 1979; Kimura & King, 1979; Li, 1980; Watterson, 1983). These models only focus on the fate of the null allele at duplicate loci, and did not calculate the preservation probability of the null allele. Subsequently, another branch of models focused on accumulation of degenerative mutations, suggesting that the preservation of duplicate copies could occur by sub-functionalization and neo-functionalization (Walsh, 1995; Force et al., 1999; Lynch & Conery 2000; Lynch, & Force, 2000; Lynch et al., 2001). Please see also Walsh's review paper for more detail on those theoretical models (Walsh, 2003). Taken together, the common features of these experimental and theoretical studies were to describe the importance of gene/genome duplication at molecular level.

The precise mechanism of fixation of duplicate loci depends on evolutionary forces, including the effective population size, mutations acting on duplicate loci, the selection coefficient for duplicate loci, and the recombination between duplicate loci. After Li (1980) and Watterson (1983) paper, the classical gene duplication model was only used in very few theoretical studies. Previous analysis of the classical models by Li and Watterson only focused on weak selection with tight or no linkage (Li, 1980; Watterson, 1983).

Here, we present a time dependent stochastic difference equation model based on Li's 1980 model to study the evolutionary dynamics of a mutant allele at duplicate loci. We investigate the fixation time of a mutant allele at duplication loci during weak and strong selection with changed recombination rates. This is done via a diffusion approach.

## 2. MODELS AND METHODS

We assume the population to be composed of a random mating diploid with an effective size of *N*. We further assume that there are two loci, which may be linked or unlinked. Moreover, we suppose two alleles: alleles *A* and *a* at the first locus, and alleles *B* and *b* at the second locus. Here *A* and *B* represent wild-type (normal function) genes and *a* and *b* are mutant (null function) genes. To simplify the model, we assume allele *A* to mutate to *a* and *B* to mutate to *b* at the same rate *u*, and that the mutation cannot be reversed. We designate the recombination rate of two alleles as r. The selection coefficient matrices are given in Table 1. We denote the frequencies of gametes *ab, aB, Ab, AB* as $x_1$, $x_2$, $x_3$, $x_4$, respectively, and the frequency of *a* and *b* as $p$ and $q$, respectively, where $p = x_1 + x_2$, $q = x_1 + x_3$.

Under the assumption that zygotes are formed by random union of gametes, the mean fitness of the population is

$$\overline{w} = 1 - 2hx_1(x_2 + x_3) - sx_1^2$$

where *h* and *s* are selection coefficients constants. We assume that in the strong selection case, we have *s*=1, *h*=1 for the double null recessive model (DNR); and in the weak selection case, we have *h*=0, *s*=1 for the haploinsufficient model (HI). The genotype frequency changes after selection, mutation and recombination (for the linked case) are

$$\begin{aligned}
dx_1/dt &= [x_1(1 - sx_1 - \overline{w}) - rD]/\overline{w} + u(x_2 + x_3), \\
dx_2/dt &= [x_2(1 - hx_1 - \overline{w}) + rD]/\overline{w} + u(x_4 - x_2), \\
dx_3/dt &= [x_3(1 - hx_1 - \overline{w}) + rD]/\overline{w} + u(x_4 - x_3), \\
dx_4/dt &= [x_4(1 - \overline{w}) - rD]/\overline{w} - 2ux_4,
\end{aligned} \qquad (1)h$$

where $D = x_2 x_3 - x_1 x_4$, which is the linkage disequilibrium constant. According to Ito's stochastic difference theory (Maruyama & Takahata, 1981; Øksendal, 2003), *x* can be considered as a *k*-dimensional diffusion process with diffusion operator

$$L = \frac{1}{2}\sum_{i,j \leq k} V_{i,j} \frac{\partial^2}{\partial x_i \partial x_j} + \sum_{j \leq k} M_j \frac{\partial}{\partial x_j}, \qquad (2)$$

where $V_{i,j}$ is diffusion coefficient and $M_j$ is the drift coefficient. Eq. *(2)* can be approximated by the following stochastic differential equation

$$d\vec{X} = \vec{V}d\vec{B} + \vec{M}dt, \qquad (3)$$

where $\vec{X} = (x_1, \cdots, x_k)^T$, $\vec{M} = (M_1, \cdots, M_k)^T$, $\vec{B}$ is a vector of mutually independent Brownian motion. From *Eq. (3)* we can get a stochastic difference equations of the time dependent gamete frequencies $x_1(t)$, $x_2(t)$, $x_3(t)$, $x_4(t)$ as:

$$\Delta x_1(t) = V_{1,1}B_1(\Delta t) + V_{1,2}B_2(\Delta t) + V_{1,3}B_3(\Delta t) + M_1\Delta t,$$

$$\Delta x_2(t) = V_{2,1}B_1(\Delta t) + V_{2,2}B_2(\Delta t) + V_{2,3}B_3(\Delta t) + M_2\Delta t, \qquad (4)$$

$$\Delta x_3(t) = V_{3,1}B_1(\Delta t) + V_{3,2}B_2(\Delta t) + V_{3,3}B_3(\Delta t) + M_3\Delta t,$$

$$\Delta x_4(t) = -\{\Delta x_1(t) + \Delta x_2(t) + \Delta x_3(t)\},$$

where the coefficient matrix V is

$$V = \begin{pmatrix} x_1(1-x_1) & -x_1x_2 & -x_1x_3 \\ -x_2x_1 & x_2(1-x_2) & -x_2x_3 \\ -x_3x_1 & -x_3x_2 & x_3(1-x_3) \end{pmatrix}$$

, and $\vec{M}$ is a non-negative definite matrix defined as

$$\vec{M} = 2N \bullet \begin{pmatrix} dx_1/dt \\ dx_2/dt \\ dx_3/dt \end{pmatrix}.$$

The Euler-Maruyama method (Higham, 2001) can be used to solve the stochastic differential equation Eq. *(4)*. During calculations, we choose a small $\Delta t$ in the difference equations Eq. *(4)* to guarantee convergence and accuracy.

## 3. RESULTS

There are four different parameters that can vary in this study: the effective population size (*N*), the mutation rate (*u*), the recombination rate(*r*) and the selection coefficient (*h, s*). We assume *r*=0.5 for free recombination; *r*=0.001 for light linkage; and *r*=0 for no linkage. We introduce *θ* as the product of the effective population size (*N*) and the mutation rate (*u*).

For each case of a given set of parameter values, simulations were repeated at least 500 times. Every path started from the same initial condition *p=q=*0 and stopped at the point where *p* or *q* was first arrived at a fixed phase. Using the evolutionary dynamics of gamete frequencies in Eq. *(4)*, we investigated how evolutionary forces influence the fixation time of a mutant allele at duplicate loci (*Table2*). According to these results the fixation time of a mutant allele depends mostly on selection, rather than neutral drift. Linkage seems to have only a small effect on the fixation of a mutant allele at duplicate loci, but may influence the evolutionary fate of duplicate genes. According to the DNR model, linkage between loci may influence the mutant allele fixation. In the HI model linkage has little effect on the fixation time of a mutant allele at duplicate genes due to the rapid evolution pressure, but it may have an influence on the fixation frequencies for *p* or *q*. Selection keeps both copies in case of a mutational pressure for silencing at duplicate loci.

We further investigated, in which way the fixation frequency was influenced, when both, p and q, were fixed initially (*Table3*). The fixation frequency of a mutant allele is influenced by selection and recombination. Tight linkage with strong selection would lead to a high fixation frequency of a mutant allele at duplicate loci. The selection coefficient *h*, rather than the selection coefficient *s*, has a dominant effect on the fixation frequencies for *p+q*. *s* is the fitness of the lethal mutant *aabb*. Increased *s* and decreased *h* increase the fixation frequency of the mutant type. In case of a tight linkage (*r*=0.5), the increase of the fixation frequency of a mutant allele have positive correlation to *θ*, while in the case of no linkage or loose linkage (*r*= 0 or 0.0001), the fixation frequency of a mutant allele is only weakly influenced by selection (fixed coefficient *h*. varied coefficient *s*). The fixation frequency of a mutant allele shows a much higher variation in a large population than in a small population. Tight linkages influence the probability of duplicate-gene preservation. In Figure 1, we show the fixation frequency of *p+q* in the case of mutation rate at *u=1.0e-6* with recombination rate, population size and selection coefficients.

Our simulation shows that deleterious mutations can never be fixed in a population and that an accumulation of mutations occurs shortly after gene duplication. This is in agreement with previous theoretical studies suggesting that tight-linkage and positive selection may increase the probability of sub-functionalization and that the copies of a duplicate gene might accumulate neutral information (Lynch et al., 2001).

It's interesting that the selection coefficient, *h*, could never be lower than 0.7, because otherwise the fixation frequencies of duplicate genes would be larger than 1, which is not allowed by the stochastic differential equation (*Eq. 4*). If *h*=0, the fixation frequencies of duplicate genes at the same time could be 1. This is in accordance to a study by Li and Watterson (Li, 1980; Watterson, 1983). This implies that selection favoring modifiers of dominance would be weak and unable to overcome a genetic drift in the population. Wright's physiological theory predicts that haploinsufficient genes should have more paralogs than haplosufficient genes because selection could increase the dosage for dosage-sensitive (haploinsufficient) genes (Wright, 1934). Clearly, our simulation is compatible with Wright's prediction. Our simulation is also in agreement with the observation of duplications in Arabidopsis thaliana, which provides evidence of positive selection (Kondrashov et al., 2002; Moore & Purugganan 2003; Moore & Purugganan 2005). Our simulation further supports the hypothesis that most duplicate genes are fixed by positive selection for increased gene dosage.

## 4. DISCUSSION

We studied the fixation processes of a mutant allele in a population via gene duplication. We here focused on the theoretical aspects of how evolutionary forces influence the evolutionary fate of duplicate genes. This was done via simulating the fixation time and fixation frequencies of a mutant allele at duplicate loci. Our simulation results demonstrated that the evolutionary trajectories and evolutionary fate of duplicate genes is a complex process that is affected by the recombination rate, the mutation rate, the effective population size and the intensity of selection. The results present here suggest that recombination and selection, rather than drift, play a key role in duplicate gene evolution.

Recently, gene duplication has been widely investigated in genomes of organism (Zhang, 2003; Moore & Purugganan, 2003; Bowers et al., 2003; Blanc & Wolfe, 2004; Clauss & Mitchell-Olds, 2004; Semon & Wolfe, 2008; Kleinjan et al., 2008). These studies showed that selection always favors duplications increasing fitness, and that recombination helps the fixation time of mutant allele at duplicate loci. Selectively neutral duplications should be very rare, for the changes in the number of genes are rare.

Our study on recombination rate and selection of duplicate loci reveals that strong selection may shorten the fixation time of null alleles and the dominance of wild-type alleles should be considered in theoretical models. Our results also show that linkage only has a minor effect on the fixation time of mutant allele at duplicate loci, but recombination with strong selection plays an important role in the fixation frequency of mutant allele at duplicate loci. Coexistent mutant alleles in an organism lead to a complex evolutionary fate of duplicate genes, which might lead to a sub-functionlization or neo-functionaliztion processes after long evolution time.

In summary our work suggests that strong selection can reduce the loss of duplicated gene and that tight linkage with other evolutionary forces may result in differential evolutionary fates. It is intriguing to ask why duplicated gene can be preserved in the genome and what causes the complexity of a genome over long evolutionary time periods. It should thus be considered in the modeling of evolutionary dynamics of gene duplication in the future. For the exact duplicate event is hard to define, it should be noted that our ideal model present here is too strict for real duplication events. Our theoretical simulation suggests that selection, rather than drift, plays an important role in the establishment of duplicate loci, and that recombination occur with strong selection helps a mutant allele to stay at duplicate loci.

I would like to thank Xue Cheng for inspire me to work on the gene duplication. I also thank Prof. Dr. Fu Yunxin, Dr. Zhai Weiwei and Dr. Liu Shuqun for many discussion and suggestion. Thanks are also due to Prof. Dr. Holger Fröhlich for editorial assistance and construction advisements on the manuscript. This work was supported by NRW State within the B-IT Research School.

**Tables**

Table1. Genotype fitness

Table2. The first fixation time of *p* or *q*

Table 3. The fixation frequencies of *p* and *q*

*h, s* are selection coefficient in Table 1, *h*=1.0 to 0.7, *s*= 0.0 to 1.0, here I only show *s*= 0.0, .6, 1.0 and *h*=1.0, 0.7.  *r* is the recombination rate, *r*= 0.5, 0.1, 0.0001, 0, I omit *r*=0.0001 for it similar to *r*= 0.  *N*=1, 2, 3, 4, 5 correspond to *N*= 10 ^ 1, 2, 3, 4, 5.  *u*= -5, -6, -7, -8 correspond to *u*= 1.0e^-5, -6, -7, -8.

*Table 1. Genotype fitnesses*

| genotypes | *AB* | *aB* | *Ab* | *ab* |
|---|---|---|---|---|
| *AB* | 1 | 1 | 1 | 1 |
| *aB* | 1 | 1 | 1 | 1-*h* |
| *Ab* | 1 | 1 | 1 | 1-*h* |
| *ab* | 1 | 1-*h* | 1-*h* | 1-*s* |

\* *h, s* are selection coefficients.

Table 2. The first fixation time of $p$ or $q$

|  |  | Strong selection case | | | | Weak selection case | | | |
|---|---|---|---|---|---|---|---|---|---|
| $r$ | $u$ | $N=1.0e+3$ | $1.0e+4$ | $1.0e+5$ | $1.0e+6$ | $N=1.0e+3$ | $1.0e+4$ | $1.0e+5$ | $1.0e+6$ |
| 0 | 1.0e-5 | 262 | 255 | 253 | 253 | 729.8 | 709.9 | 853.2 | 1654 |
|  | 1.0e-6 | 254 | 255 | 255 | 255 | 623.5 | 710.4 | 754.1 | 749.3 |
|  | 1.0e-7 | 253 | 252 | 252 | 252 | 924.4 | 628.4 | 655.3 | 6523.3 |
| 1.0e-4 | 1.0e-5 | 262 | 255 | 252 | 252 | 743.1 | 126.4 | 1306 | 2131 |
|  | 1.0e-6 | 256 | 257 | 253 | 253 | 623.4 | 714.4 | 1305 | 1553 |
|  | 1.0e-7 | 253 | 253 | 252 | 252 | 1451 | 703.6 | 704.8 | 695.7 |
| 1.0e-3 | 1.0e-5 | 262 | 255 | 252 | 252 | 769.7 | 1264 | 1549 | 2907 |
|  | 1.0e-6 | 254 | 255 | 253 | 253 | 684.7 | 723 | 706.1 | 1530 |
|  | 1.0e-7 | 253 | 253 | 252 | 252 | 1650 | 604.8 | 686.7 | 754.1 |
| 1.0e-1 | 1.0e-5 | 262 | 255 | 252 | 252 | 759.3 | 1519 | 1655 | -- |
|  | 1.0e-6 | 266 | 260 | 254 | 258 | 697.9 | 1305 | 1660 | 1639 |
|  | 1.0e-7 | 296 | 270 | 258 | 254 | 1275 | 751.2 | 1771 | -- |
| 0.5 | 1.0e-5 | 270 | 256 | 252 | 252 | 1268 | 1749 | 2321 | -- |
|  | 1.0e-6 | 255 | 265 | 257 | 254 | 817.8 | 1652 | 1569 | -- |
|  | 1.0e-7 | 320 | 294 | 267 | 268 | 1015 | 2738 | -- | -- |

\* -- the data could not be got in a same $\Delta t$ with other case.

Table 3. The fixation frequency of p and q.

| h | r | u | s=0 | | | | | s=0.6 | | | | | s=1 | | | | |
|---|---|---|---|---|---|---|---|---|---|---|---|---|---|---|---|---|---|
| | | | N=1 | 2 | 3 | 4 | 5 | N=1 | 2 | 3 | 4 | 5 | N=1 | 2 | 3 | 4 | 5 |
| 1 | 0.5 | −5 | 0.51 | 0.51 | 0.58 | 0.58 | 0.58 | 0.51 | 0.56 | 0.58 | 0.58 | 0.58 | 0.51 | 0.56 | 0.57 | 0.58 | 0.58 |
| | | −6 | 0.56 | 0.56 | 0.58 | 0.58 | 0.58 | 0.5 | 0.57 | 0.58 | 0.58 | 0.58 | 0.5 | 0.57 | 0.58 | 0.58 | 0.58 |
| | | −7 | 0.49 | 0.52 | 0.57 | 0.57 | 0.58 | 0.5 | 0.53 | 0.57 | 0.57 | 0.58 | 0.5 | 0.53 | 0.57 | 0.58 | 0.58 |
| | | −8 | 0.5 | 0.54 | 0.58 | 0.58 | 0.58 | 0.5 | 0.54 | 0.58 | 0.58 | 0.58 | 0.5 | 0.54 | 0.58 | 0.58 | 0.58 |
| | 0.1 | −5 | 0.5 | 0.53 | 0.53 | 0.53 | 0.53 | 0.5 | 0.52 | 0.53 | 0.53 | 0.53 | 0.5 | 0.53 | 0.53 | 0.53 | 0.53 |
| | | −6 | 0.5 | 0.51 | 0.51 | 0.53 | 0.53 | 0.5 | 0.51 | 0.52 | 0.53 | 0.53 | 0.5 | 0.51 | 0.52 | 0.53 | 0.53 |
| | | −7 | 0.49 | 0.5 | 0.5 | 0.52 | 0.53 | 0.5 | 0.5 | 0.51 | 0.52 | 0.53 | 0.5 | 0.5 | 0.51 | 0.52 | 0.53 |
| | | −8 | 0.5 | 0.5 | 0.5 | 0.51 | 0.52 | 0.5 | 0.5 | 0.5 | 0.51 | 0.51 | 0.5 | 0.5 | 0.5 | 0.51 | 0.53 |
| | 0 | −5 | 0.5 | 0.5 | 0.5 | 0.5 | 0.5 | 0.5 | 0.5 | 0.5 | 0.5 | 0.5 | 0.5 | 0.5 | 0.5 | 0.5 | 0.5 |
| | | −6 | 0.5 | 0.5 | 0.5 | 0.5 | 0.5 | 0.5 | 0.5 | 0.5 | 0.5 | 0.5 | 0.5 | 0.5 | 0.5 | 0.5 | 0.5 |
| | | −7 | 0.49 | 0.5 | 0.5 | 0.5 | 0.5 | 0.5 | 0.5 | 0.5 | 0.5 | 0.5 | 0.5 | 0.5 | 0.5 | 0.5 | 0.5 |
| | | −8 | 0.5 | 0.5 | 0.5 | 0.5 | 0.5 | 0.5 | 0.5 | 0.5 | 0.5 | 0.5 | 0.5 | 0.5 | 0.5 | 0.5 | 0.5 |
| 0.7 | 0.5 | −5 | 0.73 | 0.85 | 0.89 | 0.89 | 0.9 | 0.73 | 0.85 | 0.88 | 0.89 | 0.89 | 0.73 | 0.85 | 0.88 | 0.89 | 0.89 |
| | | −6 | 0.71 | 0.85 | 0.89 | 0.89 | 0.9 | 0.71 | 0.85 | 0.88 | 0.89 | 0.89 | 0.71 | 0.85 | 0.88 | 0.89 | 0.89 |
| | | −7 | 0.71 | 0.73 | 0.89 | 0.89 | 0.9 | 0.71 | 0.72 | 0.86 | 0.88 | 0.89 | 0.71 | 0.73 | 0.86 | 0.88 | 0.89 |
| | | −8 | 0.71 | 0.75 | 0.88 | 0.89 | 0.9 | 0.71 | 0.75 | 0.88 | 0.89 | 0.89 | 0.71 | 0.75 | 0.88 | 0.88 | 0.89 |
| | 0.1 | −5 | 0.71 | 0.78 | 0.78 | 0.78 | 0.77 | 0.72 | 0.77 | 0.78 | 0.77 | 0.77 | 0.72 | 0.78 | 0.78 | 0.78 | 0.77 |
| | | −6 | 0.71 | 0.77 | 0.77 | 0.77 | 0.77 | 0.71 | 0.73 | 0.77 | 0.77 | 0.77 | 0.71 | 0.73 | 0.77 | 0.77 | 0.77 |
| | | −7 | 0.71 | 0.71 | 0.74 | 0.76 | 0.77 | 0.71 | 0.71 | 0.74 | 0.76 | 0.77 | 0.71 | 0.71 | 0.74 | 0.76 | 0.77 |
| | | −8 | 0.71 | 0.71 | 0.72 | 0.74 | 0.75 | 0.71 | 0.71 | 0.72 | 0.74 | 0.75 | 0.71 | 0.71 | 0.72 | 0.74 | 0.75 |
| | 0 | −5 | 0.71 | 0.72 | 0.72 | 0.72 | 0.72 | 0.71 | 0.72 | 0.72 | 0.72 | 0.72 | 0.71 | 0.72 | 0.72 | 0.72 | 0.72 |
| | | −6 | 0.71 | 0.72 | 0.72 | 0.72 | 0.72 | 0.71 | 0.71 | 0.72 | 0.72 | 0.72 | 0.71 | 0.71 | 0.72 | 0.71 | 0.72 |
| | | −7 | 0.71 | 0.71 | 0.71 | 0.71 | 0.71 | 0.71 | 0.71 | 0.71 | 0.72 | 0.71 | 0.71 | 0.71 | 0.71 | 0.71 | 0.71 |
| | | −8 | 0.71 | 0.71 | 0.71 | 0.71 | 0.71 | 0.71 | 0.71 | 0.71 | 0.71 | 0.71 | 0.71 | 0.71 | 0.71 | 0.71 | 0.71 |

**Figures**

Figure 1. The fixation frequencies of a mutant allele at duplicate loci.

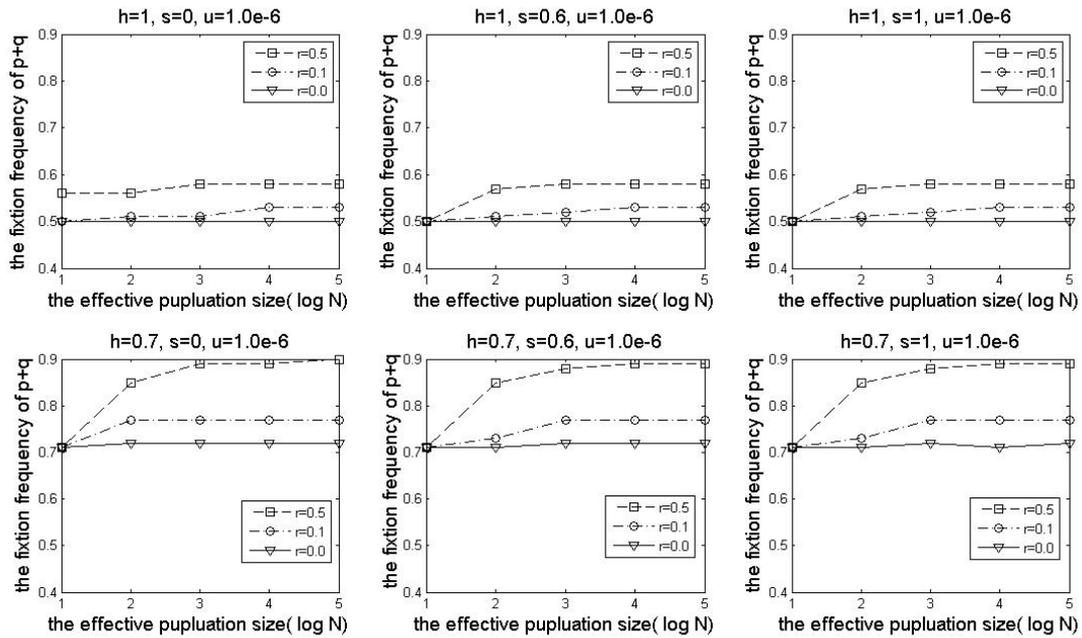

Figure 1. The fixation frequency of a mutant allele at duplicate loci. We show the fixation frequency of mutant allele (*p+q*) in the case of *u=1.0e-6* with varied recombination rates, population sizes and selection coefficients.